\documentclass[aps,preprintnumbers,amsmath,twocolumn,groupedaddress,superscriptaddress,notitlepage,nofootinbib,prd]{revtex4-1}
\usepackage[utf8]{inputenc}
\usepackage{indentfirst}
\usepackage{amssymb}
\usepackage{amsmath,latexsym}
\usepackage{graphicx}
\usepackage{float}
\usepackage[margin=1.0in]{geometry}
\usepackage[mathscr]{euscript}
\usepackage{color}
\usepackage[dvipsnames]{xcolor}
\usepackage{soul}
\usepackage{slashed}
\usepackage{feynmf}
\usepackage{subcaption}
\usepackage{braket}
\usepackage[normalem]{ulem}
\usepackage{todonotes}
\usepackage{comment}
\usepackage{makecell}

\usepackage{dsfont}

\newcommand{\beq}{\begin{equation}}
\newcommand{\eeq}{\end{equation}}
\newcommand{\bea}{\begin{eqnarray}}
\newcommand{\eea}{\end{eqnarray}}

\long\def\beqs#1\eeqs{\beq\begin{split} #1 \end{split}\eeq}

\definecolor{MyRed}{RGB}{153,0,13}
\PassOptionsToPackage{hyphens}{url}
\usepackage[colorlinks=true,backref=true,linktocpage=true,citecolor=MyRed,urlcolor=MyRed,linkcolor=MyRed,pdfpagemode=UseOutlines]{hyperref}

\hypersetup{%
  bookmarksnumbered=true,
  pdftitle = {},
  pdfsubject = {},
  pdfauthor = {},
  pdfkeywords = {}
}

\usepackage{microtype}

\usepackage{pgfplotstable,array,siunitx}
\usepackage{multirow}
\usepackage{booktabs}

\usepackage{tikz}

\preprint{INT-PUB-20-030}

\newcommand{\fig}[1]{Fig.~\ref{#1}}
\newcommand{\tab}[1]{Table~\ref{#1}}

\begin{document}
\title{A Machine Learning Approach to Trapped Many-Fermion Systems}

\author{Paulo F. Bedaque}
\email{bedaque@umd.edu}
\affiliation{Department of Physics,
University of Maryland, College Park, MD 20742}

\author{Hersh Kumar}
\email{hekumar@umd.edu}
\affiliation{Department of Physics,
University of Maryland, College Park, MD 20742}

\author{Andy Sheng}
\email{asheng@umd.edu}
\affiliation{Department of Physics,
University of Maryland, College Park, MD 20742}

\preprint{}

%\date{Aug. 2023}

\begin{abstract}
We apply a variational {\it Ansatz} based on neural networks to the problem of spin-$\tfrac{1}{2}$ fermions in a harmonic trap interacting through a short distance potential. We showed that standard machine learning techniques lead to a quick convergence to the ground state, especially in weakly coupled cases. Higher couplings can be handled efficiently by increasing the strength of interactions during ``training".

\end{abstract}

\maketitle

\section{Introduction}

\begin{table}[b!]
\centering
\begin{tabular}{|c|c|}
 \hline
\textbf{\makecell{ Unsupervised \\ Machine learning} } & \textbf{\makecell{Variational \\ Monte Carlo}}\\
 \hline
  \hline
       cost function  & energy\\
       \hline
       neural network & wave function\\
        \hline
       weights and biases & wave function parameters\\
 \hline
\end{tabular}
\caption{Analogy between unsupervised machine learning and Variational Monte Carlo.}
\label{tab:analogy}
\end{table}

In recent years, there has been a noteworthy and promising convergence between the methods of machine learning and quantum many-body physics. The central idea has been the recognition of a natural analogy between the training of neural networks and Variational Quantum Monte Carlo methods (VQMC). This connection has been explored both in the condensed matter/quantum chemistry context where the long-range Coulomb force is relevant (for a small sample see references \cite{PhysRevResearch.2.033429,doi:10.1080/23746149.2020.1797528,mezza,lili}) and in the nuclear/atomic trap case where short-range forces are of interest \cite{Gnech:2023prs,PhysRevResearch.4.043178,Kim:2023fwy,Fore:2022ljl,Nordhagen:2022jtv,PhysRevResearch.4.043178,Wang:2024ynn,Medvidovic:2024ihh,boson_paper,Keeble:2023rre}. 
Common uses of machine learning techniques are based on creating an artificial neural network with some specific architecture and choosing the parameters of this network in order to accomplish some task of interest. In the unsupervised learning approach,   the network parameters are tuned in order to minimize a ``cost function" --- a function of the network parameters whose value correlates with the skill in that particular task. By numerically minimizing the cost function by varying the network parameters (``training" the network) one arrives at a network with some amount of skill in that task. Similarly, in the VQMC method, one parameterizes the ground state of a system and minimizes the energy in relation to these parameters. The wave function with the smallest energy within that parameterized family is the best estimate of the ground state (it is actually a rigorous upper bound on the exact value). The analogy then is that the energy plays the role of cost function and the network is the family of parameterized wave functions (see \tab{tab:analogy}). \\

The ``Monte Carlo" part of VQMC refers to the fact that the energies and their gradients cannot be computed analytically for many-particle systems with complicated wave functions. Instead, they are estimated stochastically by Monte Carlo methods. The stochastic noise present in the final evaluation of the energy is kept to a minimum by using a long Monte Carlo chain; the noise in the gradient does not need to be very small. In fact, some stochastic noise is useful in order to avoid getting trapped in local minima of the energy during training \cite{d8d62392-9a37-31e7-ad3b-37a6f6ee8ef6}.  The great advantage of framing VQMC in the machine learning language is to use a number of tricks and techniques used in that field so that training large systems, with wave functions parameterized by many (sometimes millions of) parameters, can be readily accomplished on standard laptops. \\

In this work, we focus on a class of problems permeating several fields of Physics: systems comprised of multiple fermions interacting through short-range potentials. These systems, while conceptually simple, pose significant computational challenges due to the exponential growth of the Hilbert space with particle number and a serious sign problem in some Monte Carlo approaches. Our method leverages the aforementioned analogy to develop novel computational strategies for tackling these many-body problems. A key feature of our methodology is its ability to transfer knowledge learned about lower coupling strengths to accelerate calculations in higher coupling strength regimes. This transfer learning approach not only promises to enhance computational efficiency but also offers insights into the scaling behavior of quantum systems. \\

This paper is organized as follows: in section \ref{sec:model} we explicitly describe our model system; in section \ref{sec:methods} we review the VQMC approach and describe our neural network variational ansatz. We then present numerical results in section \ref{sec:results} and discuss future directions in section \ref{sec:discussion}.

\section{Model \label{sec:model}}

The model which we consider consists of $N_{\uparrow}$ spin-$1/2$ fermions in the up-spin state and $N_{\downarrow}$ fermions in the down-spin state in one dimension. All particles have mass $m$ and are confined by a harmonic trap. Pairs of particles in opposing spin states interact via a contact potential with coupling strength $g$. The Hamiltonian is given by 
\begin{equation}
    \hat{H} = \sum_{i=1}^N \left(-\frac{\hbar^2}{2m}\frac{\partial^2}{\partial x_i^2} + \frac{1}{2} m\omega^2 x_i^2\right) + \sum_{i<j}^N g \delta(x_i - x_j), \label{model}
\end{equation}
where $N = N_{\uparrow} + N_{\downarrow}$. Note that the contact interaction does not affect pairs of particles in the same spin state, since the particle exchange anti-symmetry implies that the wave function vanishes as $x_i - x_j \rightarrow 0$. We use units in which $m = \omega = \hbar = 1$. \\

In the case of two fermions, one in each spin state, the model is analytically solvable \cite{busch, WILSON20141065}. For systems with three or more interacting fermions, no known analytic solutions exist for the ground state; however, numerical calculations have been performed up to $\sim10$ particles \cite{levinsen2015strong, volosniev2014strongly, sowinski2013few, PhysRevA.88.021602}. The gas of $N_{\uparrow}$ trapped fermions with a single impurity ($N_{\downarrow} = 1$) has also been examined using a diffusion Monte Carlo (DMC) method \cite{PhysRevA.88.021602}. An approximate expression for the ground state energy of the impurity system as the number of spin-up fermions increases was obtained by modifying McGuire's homogeneous case ($\omega = 0$) expression \cite{mcguire1965interacting}. The modified McGuire expression 
\begin{equation}
\begin{split}
    &E_{\text{imp}} = \frac{N_{\uparrow}^2 + 1}{2} \\
    &+ \frac{N_{\uparrow}\gamma_t}{\pi^2}\left[1 - \frac{\gamma_t}{4} + \left(\frac{\gamma_t}{2\pi} + \frac{2\pi}{\gamma_t}\right)\arctan \frac{\gamma_t}{2\pi}\right]
\end{split} \label{mcguire}
\end{equation}
where $\gamma_t = \frac{\pi g}{2}\sqrt{\frac{2}{N_{\uparrow}}}$ was found to approximate the DMC results closely at large $N_{\uparrow}$ for both weak and strong coupling strengths \cite{PhysRevA.88.021602}.

\section{Methods\label{sec:methods}}

\subsection{VQMC}

To search for ground state solutions of the many-fermion systems, we employ the variational principle --- 
\begin{equation}
    \langle E(\theta)\rangle = \frac{\int dX\:\Psi(X, \theta)\hat{H}\Psi(X, \theta)}{\int dX\:\Psi(X, \theta)^2} \geq E_0,
\end{equation} where $X$ stands for the coordinates of the system (the $N$ particle coordinates in our case) and $\theta$ stands for the set of wave function parameters.
By parameterizing a suitable \textit{Ansatz} function, an approximation for the ground state wave function may be obtained by minimizing $\langle E(\theta)\rangle$ with respect to the parameters $\theta$ \footnote{The ground state wave function for a non-degenerate many-body, time-independent quantum system in one dimension can be taken to be strictly real.}. \\

The VQMC technique involves computing the energy $\langle E(\theta)\rangle$ and its gradient $\partial \langle E(\theta)\rangle / \partial \theta$ numerically via Monte Carlo sampling. The energy and gradient can be written as
\begin{equation}
\begin{split}
    \langle E\rangle &= \frac{\int dX\:\Psi(X)^2 \left(\Psi(X)^{-1}\hat{H}\Psi(X)\right)}{\int dX\:\Psi(X)^2} \\
    &= \langle \Psi(X)^{-1}\hat{H}\Psi(X)\rangle_{\pi} \label{VQMC}
\end{split}
\end{equation}
\begin{equation}
\begin{split}
    \frac{\partial \langle E\rangle}{\partial \theta} &= \frac{2\int dX\: \Psi(X)^2 \left[\partial_{\theta}\log(\Psi(X))(\frac{\hat{H}\Psi(X)}{\Psi(X)} - \langle E\rangle)\right]}{\int dX\:\Psi(X)^2} \\
    &= 2\langle \partial_{\theta}\log(\Psi(X))(\Psi(X)^{-1}\hat{H}\Psi(X) - \langle E\rangle) \rangle_{\pi}
\end{split}
\end{equation}
where $\langle ... \rangle_{\pi}$ denotes an expectation value over samples from the distribution 
\begin{equation}
    \pi(X,\theta) = \frac{\Psi(X, \theta)^2}{\int dX\:\Psi(X, \theta)^2}.
\end{equation}
We use the Metropolis-Hastings algorithm \cite{10.1063/1.1699114, hastings1970monte} to sample position configurations $X$ from $\pi(X, \theta)$. \\

The procedure then, is the following. We first obtain Monte Carlo samples of $X$ from $\pi(X, \theta)$ and utilize those samples to estimate $\langle E(\theta)\rangle$ and $\partial \langle E\rangle / \partial \theta$; the energy is thought of as the cost function in the machine-learning analogy and the number of samples used in its estimate is akin to the batch size. Unlike in typical machine learning applications, increasing the number of samples simply increases the precision at which the cost function is estimated and does not change its value; thus, the batch size in this context may be changed during the training procedure. In fact, increasing the ``batch size" as training progresses results in more accurate determinations of system ground states. Then, we update the wave function \textit{Ansatz} by performing a gradient descent step in parameter space towards minimizing the expectation of the energy. This process is repeated until the energy ceases to decrease within some certain error tolerance. \\

Note that the integral expressions for $\langle E\rangle$ and $\partial \langle E\rangle / \partial \theta$ involve non-zero expectation values of $\delta$-functions $\langle g\delta(x_i - x_j)\rangle_{\pi}$. For such integrals, the Markov Chain Monte Carlo method runs into a severe overlap problem in which few (if any) of the samples from $\pi(X, \theta)$ contribute to the evaluation of the integral, enormously increasing the variance in the estimate. To solve this problem, we employ a reorganization method introduced in the Appendix of Ref. \cite{boson_paper}.\\

The success of a variational calculation hinges on having a good \textit{Ansatz}.  If the true ground state wave function cannot be reached (or well approximated) by any element in the {\it Ansatz} family, the variational approach will not provide an accurate estimate for the ground state energy and properties of the system. Traditionally, insight into the Physics of the system was required since the number of parameters had to be small (for instance, the BCS \cite{bcs} ansatz has one single parameter). Numerical calculations and the Monte Carlo method allows for a larger number of parameters and more detailed wave functions. Our approach is to pursue this method to its logical conclusion, increase the  number of parameters in a systematic way in order to guarantee convergence to the right result. This approach is based on the fact that feed-forward neural networks (also known as multi-layer perceptrons) and our \textit{Ansatz} are universal function approximators \cite{HORNIK1989359, HORNIK1991251} and our \textit{Ansatz} in the limit of a large number of nodes/layers. Therefore, given a network with enough nodes, we expect that we can explore a comprehensive functional space for the ground state without needing to change the architecture of the \textit{Ansatz}. \textit{After} the ground state is found, one can try to examine the optimal parameters (or the most relevant) and extract physical insight \cite{samek2017explainable}. At this time this step still poses challenges and will be left for a further publication.

\subsection{Wave Function Ansätze}

An \textit{Ansatz} for the ground state wave function for a system of $N_\uparrow$ spin up fermions and $N_\downarrow$ spin down fermions should enforce anti-symmetry under the exchange of positions of fermions of the same spin. For a system of $N = N_\uparrow + N_\downarrow$ fermions, 
\begin{equation}
    \begin{split}
        &\Psi(..., x_i, ...., x_j, ...) = -\Psi(..., x_j, ..., x_i, ...),\qquad\qquad \\
        &\qquad\qquad\text{for} \: 1 \leq i \neq j \leq N_{\uparrow}\:\text{or}\: N_{\uparrow} + 1 \leq i \neq j \leq N \label{symmetry}
    \end{split}
\end{equation}
where $\{X^{\uparrow}\} = x_1,\dots,x_{N_\uparrow}$ denote the positions of the spin up fermions, and $\{X^{\downarrow}\} = x_{N_\uparrow + 1},\dots, x_{N}$ denote the positions of the spin down fermions. \\

We construct our wave function \textit{Ansatz} as
\begin{equation}
    \Psi(X, \theta) = \Phi^{\uparrow}(X, \theta)\Phi^{\downarrow}(X, \theta)e^{-\sum_{i = 1}^{N}x_i^2}.
    \label{ansatz}
\end{equation}
The  parameterized functions $\Phi^{\uparrow}$ and $\Phi^{\downarrow}$ capture the fermion-fermion interactions in the model, including those due to the Pauli principle. $\Phi^{\uparrow}$ is constructed such that it is anti-symmetric under the exchange of the coordinates of spin-up fermions $\{X^{\uparrow}\}$ and \textit{symmetric} under exchange of the coordinates of spin-down fermions $\{X^{\downarrow}\}$. $\Phi^{\downarrow}$, on the other hand, is \textit{symmetric} under the exchange of spin-up positions and anti-symmetric under the exchange of spin-down positions. The Gaussian factor in the \textit{Ansatz} ensures that the wave function vanishes at infinity. That way the integrals in Eq. \ref{VQMC} converge regardless of the values the neural networks' parameters take during the training process. The product of these three functions ensure that $\Psi(X, \theta)$ satisfies the required exchange properties in Eq. \ref{symmetry}. We now discuss how to construct $\Phi^{\uparrow}$ and $\Phi^{\downarrow}$.\\

Any function $f(X)$ which is anti-symmetric under the exchange of a subset of its coordinates $\{X^{a}\} = x_1,...,x_m$ and symmetric under the exchange of the remaining coordinates $\{X^{s}\} = x_{m+1},...,x_M$ can be written as a single of Slater determinant \cite{Pfau_2020} of $m$ functions
\begin{equation}
    f(X) = \begin{vmatrix}
        \phi_{1,1} & \phi_{1,2} & \dots & \phi_{1,m}\\
        \phi_{2,1} & \phi_{2,2} & & \\
        \vdots & & \ddots & \\
        \phi_{m,1} & & & \phi_{m,m}
    \end{vmatrix}
\end{equation}
where the $\phi_{i, j} = \phi_i(x_j, \{X^{a/j}\}, \{X^{s}\})$ are symmetric under exchanges of coordinates among $\{X^{a/j}\}$ and among $\{X^{s}\}$. Here $\{X^{a/j}\}$ denotes the set of particle positions $\{X^{a}\}$ \textit{excluding} $x_j$ (with $1 \leq j \leq m$). No symmetries are enforced for an exchange of coordinates between sets $\{X^{a/j}\}$ and $\{X^s\}$. \\

To symmetrize over the sets of coordinates $\{X^{a/j}\}$ and $\{X^s\}$, we map the coordinates into two sets of variables $\{\xi^{a/j}\}$ and $\{\xi^{s}\}$ given by
\begin{equation}
    \begin{split}
        &\xi^{a/j}_{k} = \sum_{i \neq j}^{m} \left(\frac{x_i}{w_a}\right)^k, \\
        &\xi^{s}_{k} = \sum_{i = m+1}^{M} \left(\frac{x_i}{w_s}\right)^k.
    \end{split}
\end{equation}
The number of elements in $\{X\}$ and $\{\xi\}$ are the same. The scaling of the coordinates by constants $w_a$ and $w_s$  ensures numerical stability by preventing  the inputs to $\phi_{i}$ from growing too large. The symmetric polynomials $\xi$ above are chosen specially such that given the set of $\{\xi\}$ the original set of coordinates $\{X\}$, \textit{but not their ordering}, can be uniquely recovered \cite{boson_paper}. By feeding in $\{\xi^{a/j}\}$ and $\{\xi^s\}$ as inputs into the $\phi_i$,
\begin{equation}
    \phi_i(x_j, \{X^{a/j}\},\{X^s\}) \rightarrow \phi_i(x_j, \{\xi^{a/j}\},\{\xi^s\})
\end{equation}
the $\phi_i$ satisfy the required symmetry properties without loss of information about the original coordinates. 
For example, for $m = 3$ and $M = 5$, we have for $j = 2$:
\begin{equation}
\begin{split}
    &\phi_{i, 2} = \phi_i(x_2, \{X^{a/2}\}, \{X^s\}) \\
    &= \phi_i\left(x_2, \frac{x_1+x_3}{w_a}, \frac{x_1^2+x_3^2}{w_a^2}, \frac{x_4+x_5}{w_s}, \frac{x_4^2+x_5^2}{w_s^2}\right)
\end{split}
\end{equation}

Therefore to write general functions $\Phi^{\uparrow}$ and $\Phi^{\downarrow}$ which satisfy their respective coordinate anti-symmetry and symmetry requirements, we have
\begin{equation}
    \Phi^{\uparrow}(X) = \frac{1}{C}\begin{vmatrix}
        \phi^{\uparrow}_{1,1} & \phi^{\uparrow}_{1,2} & \dots & \phi^{\uparrow}_{1,N_{\uparrow}}\\
        \phi^{\uparrow}_{2,1} & \phi^{\uparrow}_{2,2} & & \\
        \vdots & & \ddots & \\
        \phi^{\uparrow}_{N_{\uparrow},1} & & & \phi^{\uparrow}_{N_{\uparrow},N_{\uparrow}}
    \end{vmatrix}
\end{equation}
where $\phi^{\uparrow}_{i,j} = \phi^{\uparrow}_i(x_j, \{\xi^{\uparrow/j}\}, \{\xi^{\downarrow}\})$ (with $1 \leq j \leq N_{\uparrow}$ the position of a spin-up fermion) and 
\begin{equation}
    \Phi^{\downarrow}(X) = \frac{1}{C}\begin{vmatrix}
        \phi^{\downarrow}_{1,N_{\uparrow} + 1} & \phi^{\downarrow}_{1,N_{\uparrow} + 2} & \dots & \phi^{\downarrow}_{1,N}\\
        \phi^{\downarrow}_{2,N_{\uparrow} + 1} & \phi^{\downarrow}_{2,N_{\uparrow} + 2} & & \\
        \vdots & & \ddots & \\
        \phi^{\downarrow}_{N_{\downarrow},N_{\uparrow} + 1} & & & \phi^{\downarrow}_{N_{\downarrow},N}
    \end{vmatrix}
\end{equation}
where $\phi^{\downarrow}_{m, n} = \phi^{\downarrow}_{m}(x_n, \{\xi^{\uparrow}\}, \{\xi^{\downarrow/n}\})$ (with $N_{\uparrow}+1 \leq n \leq N$ the position of a spin-down fermion). The constant $C$ is chosen to ensure that $\Phi^{\uparrow}$ and $\Phi^{\downarrow}$ remain of reasonable magnitude as the number of particles in the system is increased. \\

We parameterize each of the $N = N_{\uparrow} + N_{\downarrow}$ functions $\phi^{\uparrow}_{i}$ and $\phi^{\downarrow}_{m}$ by a unique feed-forward neural network. The parameters $\theta$ for the wave function \textit{Ansatz} are the $N$ sets of weights and biases of the $\phi$-function neural networks. As the number of nodes for the $\phi$'s is increased, the \textit{Ansatz} in Eq. \ref{ansatz} can be trained towards representing the ground state of any system by searching through a general space of functions satisfying the correct fermionic exchange anti-symmetries. 

\subsection{Neural Network Architecture}

In the limit of large neural networks representing the functions $\phi^{\uparrow}_i$ and $\phi^{\downarrow}_m$, our \textit{Ansatz} can represent any wave function. In practice, there is a tension between  having large networks that can represent a large class of functions (including those close to the ground state) and the computational cost of training these networks. Unfortunately, there is no general rule to determine the optimal size of the network in  any given application \cite{doi:10.1080/01431160802549278}. What can be done is to systematically increase the network size and verify that the ground state energy stops varying beyond a certain size. Given the same number of variational parameters $\beta$, whether deep and narrow networks or shallow and wide networks more efficiently capture the ground states of interacting systems is yet to be systematically explored. We arbitrarily choose neural network structures which take $N$ inputs, go through two hidden layers of $\alpha$ nodes each, and return a single output.\\

In a feed-forward neural network, the outputs of a layer $i$
\begin{equation}
    O_i = f(\mathbf{W}I_i + \vec{b}) = I_{i+1}
\end{equation}
become the inputs for the next, $(i+1)$th layer. The function $f$ is a non-linear activation function which we apply to each layer except the output layer. We choose CELU as the activation function.
\begin{align}
    \text{CELU}(x) &= \begin{cases}
        x & x > 0 \\
        e^x - 1 & x \leq 0
    \end{cases}
\end{align}
The weights ($\mathbf{W}$) and biases ($\vec{b}$) of the neural networks are initialized randomly from a Gaussian distribution with variance $2/(n_{in}+n_{out})$, where $n_{in}$ ($n_{out}$) are the number of nodes in the layers connected by $\mathbf{W}$ \cite{pmlr-v9-glorot10a}. \\ 

Using Monte Carlo estimates of gradients $\partial \langle E\rangle / \partial \theta$, we utilize the Adam optimizer \cite{Kingma:2014vow} to perform gradient descent on the neural networks' parameters to approach the ground state. We use learning rates between $10^{-3}$ and $10^{-5}$, decreasing the learning rate as the optimizer converges closer to the ground state. The standard machine learning technique of backpropagation allows for efficient calculations of gradients. All computations were performed using the JAX software package for Python \cite{jax2018github}. JAX's just-in-time compilation, vectorization, and parallelization features make it an appealing package for neural network computations. \\

\begin{figure}[t!]
\begin{minipage}{.49\textwidth}
\includegraphics[width=1.0\textwidth]{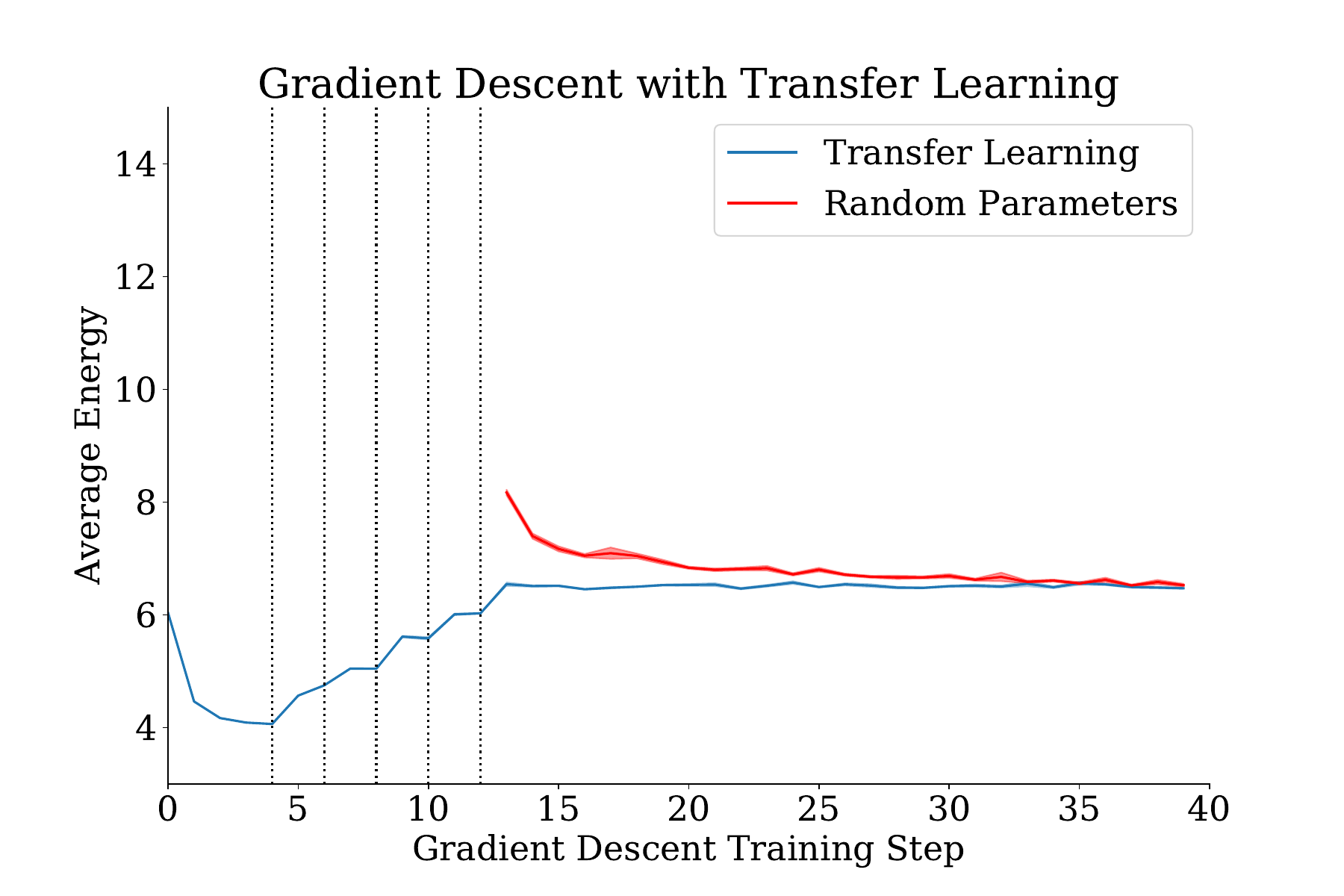}
\caption{\small{An example of transfer learning applied to a system with four fermions, $N_\uparrow = N_\downarrow= 2$, starting at $g = 0$ and with $\delta g = 0.5$, up to $g=2.5$. The vertical dotted lines denote points in the training where $g$ is increased by $\delta g$. The initial parameters for the $g=0$ starting point are chosen randomly. The red curve is a gradient descent optimization for $g=2.5$, starting with random parameters. The transfer learning optimization converges within 15 gradient descent steps, whereas the random parameter optimization converges after approximately 30 steps. \label{fig:transfer_learning_demo}}}
\end{minipage}\hspace*{\fill}
\end{figure}

\subsection{Transfer Learning}

In order to decrease the number of gradient descent steps necessary to converge to the ground state energy of a particular system, we utilize the established machine learning technique of transfer learning. In transfer learning, a machine learning model trained on one domain is used as a starting point to improve the learning of a model in a different, similar domain \cite{pan2009survey}.\\

We expect that for a system with a fixed particle count, ground state wave functions and consequently the optimal neural network  parameters are similar for interaction strengths that are near one another. Therefore, as we explore systems with different values of $g$, we employ transfer learning. After determining the set of variational parameters that approximates the ground state wave function of a system with interaction strength $g$, we use those trained neural networks' weights and biases as the set of initial parameters for the gradient descent optimization for a system with $g + \delta g$. Figure \ref{fig:transfer_learning_demo} shows that this approach allows us to save computation time and converge to the ground state of the system with $g + \delta g$ in fewer training steps (at the same learning rate) than starting again with a randomly initialized set of parameters. \\

\begin{figure}[t!]
\begin{minipage}{.48\textwidth}
\includegraphics[width=1.0\textwidth]
{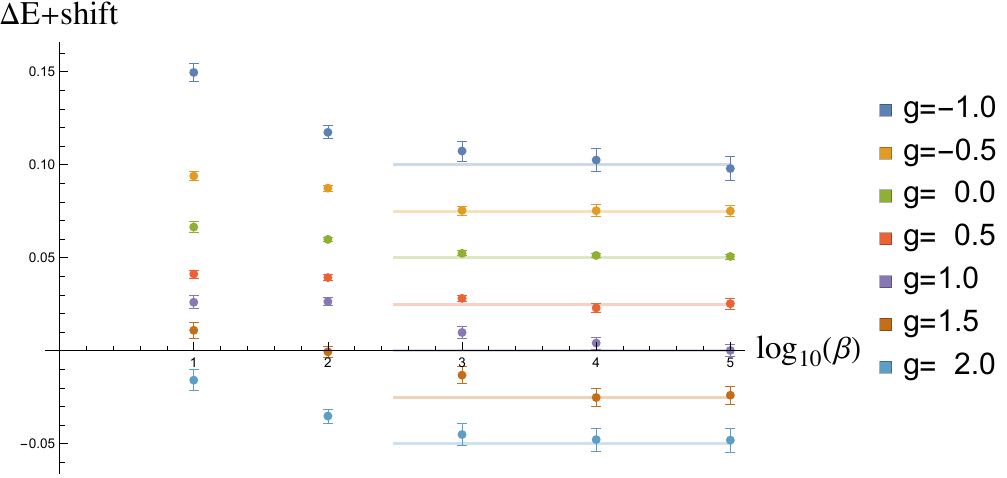}
\includegraphics[width=1.0\textwidth]
{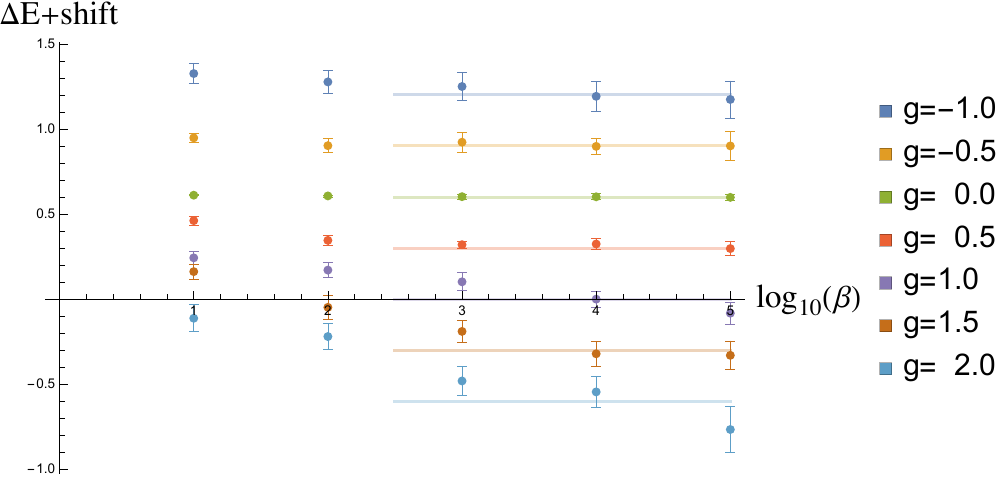}
\caption{
\small{Convergence of the ground state energy as the number of neural network parameters is increased. The horizontal lines indicate the exact solution (in (a)) and the extrapolation to large $\beta$ (in (b)). In both plots the results for each value of $g$ are shifted vertically by a different amount for clarity.
(a) The two-fermion system $N_{\uparrow} = N_{\downarrow} = 1$. The colored points are the differences of ground state energies computed with the neural network \textit{Ansatz} and the exact Busch solution. The statistical errors from the Monte Carlo calculations on some points are too small to be seen.  The horizontal lines show the analytic result. (b) The $N_{\uparrow} = N_{\downarrow} = 5$ system.
These plots show that sub-percent precision is obtained with about $10^3$ to $2\times10^3$ parameters.
\label{fig:conv}}}
\end{minipage}\hspace*{\fill}
\end{figure}

In practice, we begin with a randomly initialized set of weights and biases. Upon training those parameters to find the ground state of the non-interacting system $g = 0$, we repeatedly increase (or decrease) $g$ by $\delta g = 0.5$ and continue training --- \textit{without} re-initializing the parameters. The model's starting point for the training for the system with $g + \delta g$ is the fully trained model for the system with coupling strength $g$. That guarantees quick convergence of the ground state energy for higher and higher values of $g$. It is worth noting that the advantage of starting this process from $g=0$ relies on the fact that the {\it Ansatz} in Eq. \ref{ansatz}  converges very quickly to the non-interacting wave function starting from the initial conditions stated above. The explicit form of Eq. \ref{ansatz} does not make this apparent but it is something that we observed empirically and motivated the use of Eq. \ref{ansatz}.

\begin{figure}[b!]
\begin{minipage}{.48\textwidth}
\includegraphics[width=1.0\textwidth]{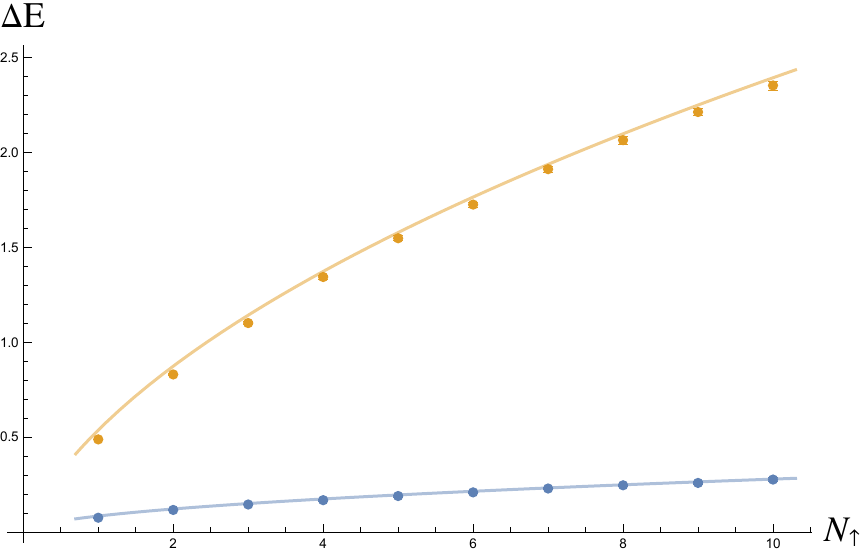}
\caption{\small{Shows the difference $\Delta E$ between the ground state energy of the interacting system $E_0$ and the ground state energy of the non-interacting, harmonic trap system $E_{\text{nonint}} = ({N_{\uparrow}}^2 + 1)/2$ as a function of $N_{\uparrow}$. The points show the neural network calculations while the solid lines plot the modified McGuire expression. The blue shows the case with interaction strength $g = 0.2$ while the yellow shows $g = 2.0$.} \label{fig:impurity}}
\end{minipage}\hspace*{\fill}
\end{figure}

\section{Results\label{sec:results}}

In order to demonstrate the flexibility of our neural network variational \textit{Ansatz} (Eq. \ref{ansatz}) in approaching the ground states of different many-body systems, we use the \textit{Ansatz} to compute the various fermionic systems described by the Hamiltonian in Sec. \ref{sec:model}.

\begin{figure}[b!]
\begin{minipage}{.48\textwidth}
\includegraphics[width=1.0\textwidth]{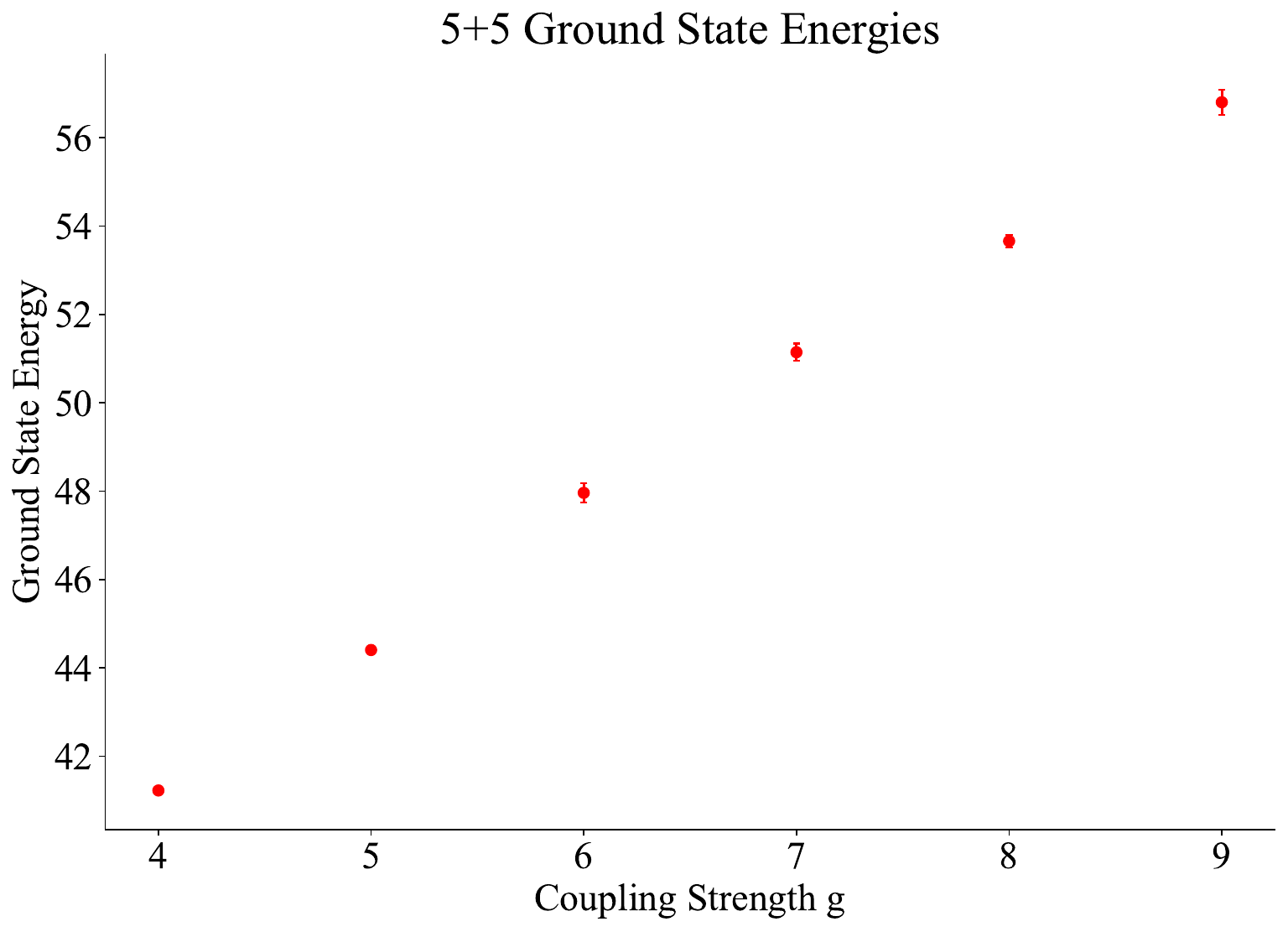}
\caption{\small{Ground state energy for a system with $N_\uparrow = N_\downarrow = 5$ for large values of $g$ computed with the neural-network-based VQMC method. \label{fig:5+5_nonperturbative}}}
\end{minipage}\hspace*{\fill}
\end{figure}

\subsection{Trapped Fermi Gas With a Single Impurity}

We first study the exactly solvable case where $N_{\uparrow} = N_{\downarrow} = 1$, with coupling strength ranging from $-1 \leq g \leq 2$ (see Fig. \ref{fig:conv}a). \\

We first analyze the system of $N_{\uparrow}$ spin up fermions and a single ($N_{\downarrow}$ = 1) spin down fermion. We compute the ground state energies of the system with the neural network \textit{Ansatz} at interaction strengths $g = 0.2$ and $g = 2.0$ for $N_{\uparrow}$ between one and ten. Astrakharchik and Brouzos \cite{PhysRevA.88.021602} show that the modified McGuire expression in Eq. \ref{mcguire} approximates the exact ground state energy of the impurity system well as an upper bound. The modified McGuire approximation becomes more accurate as $N_{\uparrow} \gg 1$. Using neural network structures of $\alpha = 15, 25, 50, 100, 150$, we find that as we increase $\alpha$ (the number of nodes in each hidden layer) and subsequently $\beta$ (the total number of \textit{Ansatz} parameters), the computed ground state energy converges to the Busch \cite{busch} solution within statistical errors. We find that $\alpha = 100$ --- yielding an \textit{Ansatz} with $\beta \approx 2\times 10^4$ --- was sufficient to capture the analytic ground state energy to sub-percent accuracy across all coupling strengths considered (see Fig. \ref{fig:conv}a). In Fig. ~\ref{fig:impurity}, we plot the Variational Monte Carlo results using our neural network \textit{Ansatz} against the modified McGuire ground state energy. We find that in both cases of $g = 0.2$ and $g = 2.0$, the neural network solution correctly captures the analytic Busch solution for $N_{\uparrow} = 1$ and closely follows the modified McGuire expression. The neural network results resemble the behavior of Ref. \cite{PhysRevA.88.021602}'s DMC results in that the numerical solutions lie below and approach the modified McGuire expression as $N_{\uparrow} \rightarrow \infty$.

\subsection{Equal Spin Populations}
We also consider the system of $N_\uparrow = N_\downarrow = 5$ fermions (see Fig.~\ref{fig:5+5_nonperturbative}). This system does not possess an analytic solution; thus, we systematically increase $\alpha$ until the estimate for the ground state energy converges. (see Fig. \ref{fig:conv}b). Neural networks with $\alpha = 150$, $\beta \approx 2\times 10^5$ successfully obtain ground state energies within sub-percent in central value to the energies calculated using the largest value of $\alpha = 200$. Thus, all results in this section were computed using neural network structures with $\alpha = 150$. \\ 

\begin{figure}[t!]
\begin{minipage}{.48\textwidth}
\includegraphics[width=1.0\textwidth]{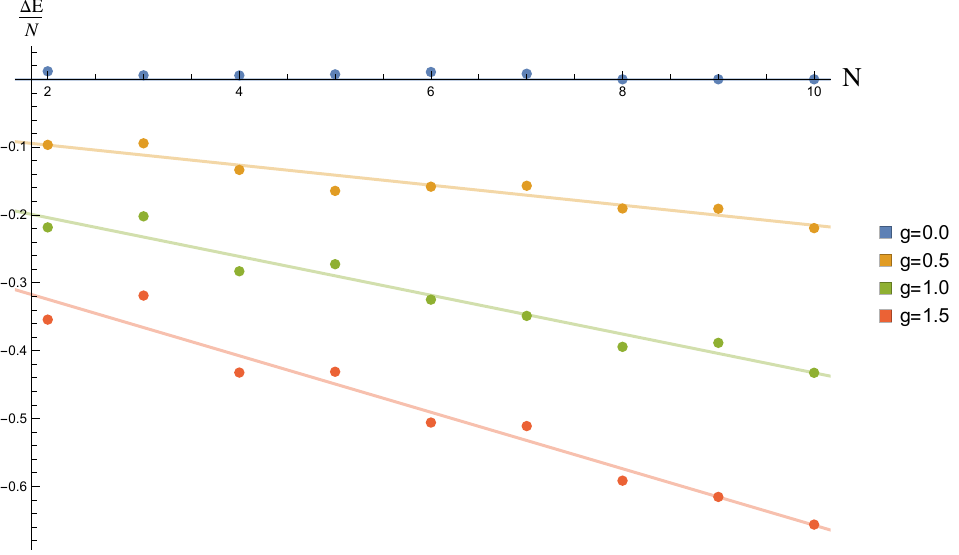}
\caption{\small{Average {\it interaction} energy per particle for systems with $N=2$ to $N=10$ particles, in the case of an attractive contact potential. For even $N$, $N_\uparrow = N_\downarrow = N/2$, and for odd $N$, one species of fermion has 1 more particle than the other. The lines are rough linear fits to make the pairing effect more pronounced as the $\delta$-function interaction becomes more attractive. The Monte Carlo error bars are too small to be seen.\label{fig:pairing}}}
\label{fig:gap}
\end{minipage}\hspace*{\fill}
\end{figure}

In \fig{fig:gap} we show the ground state energies per particle for systems
with $N_{\uparrow}=N_{\downarrow}$ and $N_{\uparrow}=N_{\downarrow}+1$ for several values of $N_{\uparrow}$. The shell effect due to the occupation of new harmonic oscillator levels is very visible but a simple model where the binding energy is accounted for by a single pairing energy multiplied by the number of available pairs -- in the manner pairing is introduced in nuclear mass formulae -- does {\it not} explain our results.

\section{Discussion \label{sec:discussion}}
We used a variational {\it Ansatz} constructed out of feed-forward neural networks for the ground state of a system of harmonically trapped fermions with a contact interaction. We demonstrated that our ansatz converges very quickly to the non-interacting ground state and that the minimization for a certain value of the interaction can be profitably used as an initial condition for a larger coupling. That way the minimization for any strength of the fermion interaction can be accomplished quickly --- in a fraction of the time required for the minimization starting from a random initial condition. We checked our results against the solvable problem with only a pair of fermions and benchmarked our approach against previous results in the literature. \\

Our results pave the way to many extensions. Three-dimensional systems and systems with a larger number of particles are obvious directions. There is no obvious impediment of performing these calculations with the technology described in this paper. However, our future goals involve  dissecting the ground states we have found in search of even simpler, but equally precise {\it Ansatze} that can then be used in more challenging problems (larger $N$, for instance) without incurring increased computational cost.
Perhaps surprisingly, the number of parameters required for a given relative precision depends weakly on the number of particles, at least in the range of $N$ we studied.

\bibliography{refs}
\end{document}